\documentstyle[twocolumn,pra,aps,eqsecnum,epsf]{revtex}
\begin{document}
\title{Necessary and Sufficient Classicality Conditions on
Photon Number Distributions}
\author{R. Simon~\cite{email1}}
\address{Institute of Mathematical Sciences, CIT Campus,
Madras 600 113\\ India}
\author{Mary Selvadoray}
\address{Institute of Mathematical Sciences, CIT Campus,
Madras 600 113\\ India}
\author{Arvind~\cite{email2}}
\address{Department of Physics, 
Indian Institute of Science, Bangalore 560 012\\ India}
\author{N. Mukunda~\cite{jncasr}}
\address{Department of Physics and Center for Theoretical Studies,\\
Indian Institute of Science, Bangalore 560 012 \\India}
\maketitle

\begin{abstract}
We exploit classical results on the Stieltjes moment
problem to obtain completely explicit necessary and
sufficient conditions for the photon number distribution of
a radiation field mode to be classical.  These conditions
are given in two forms - respectively local and global in
the individual photon number probabilities.  Equivalence of
the two approaches is demonstrated.  Detailed quantitative
statements on oscillations in the photon number
probabilities are also presented.
\end{abstract} 

\section{Introduction}
The study of nonclassical aspects of radiation from a
variety of viewpoints is of considerable and continuing
interest.  On the conceptual side one can distinguish
between phase sensitive signatures of nonclassicality and
phase insensitive ones.  The most familiar of the former is
quadrature
squeezing~\cite{slusher-prl-1985+,yuen-pra-1976+}, while in
the latter we have amplitude squeezing or subpoissonian
photon statistics~\cite{short-prl-1983+,teich-progopt}.
Higher order squeezing~\cite{hong-prl-1985+} as well as
conditions on the factorial moments~\cite{agarwal-pra-1992}
of the photon number distribution (PND) have also been
presented as sufficient conditions for nonclassicality.  In
some cases, the single mode treatments have been
generalised to the two or general multimode
situations~\cite{milburn-jphysa-1984+}.

From another point of view, a finer classification of
states of quantised radiation according to increasing
nonclassicality has been given, stressing the connection to
the specific classes of observables being
measured~\cite{arvind-characterisation}.  The case of
Gaussian-Wigner distributions for a single mode field has
been analysed in detail to give a concrete illustration of
these ideas~\cite{arvind-wigner}.

In the present paper we give a complete treatment of phase
insensitive measurements on the single mode quantized
radiation field, and develop {\em necessary and sufficient}
conditions for nonclassicality of the field.  Our focus is
on the PND, for any given state, which contains all
information concerning all phase insensitive measurements.
There are several previous partial results in this
direction.  The best known is the Mandel $Q\/$-parameter
criterion for distinguishing globally between sub and super
poissonian statistics.  This is {\em global} in the sense
that the evaluation or measurement of the parameter $Q\/$
in any given case requires knowledge of the probability for
finding $n\/$ photons in the state, {\em for all $n\/$}.
We may also mention the important generalisation of the
Mandel $Q\/$ parameter achieved by Agarwal and
Tara~\cite{agarwal-pra-1992}.  Our principal motivation is
to look for {\em local} conditions on the PND, involving
only a {\em finite} number of photon probabilities, which
will serve as signatures to distinguish between classical
and nonclassical states.  This approach or viewpoint allows
us to give a rather detailed analysis of
oscillations~\cite{schleich-nature-1987+,agarwal-pra-1989+}
in the PND, and to see precisely which kinds of
oscillations stem from nonclassicality and which do not.
As for the fact that we are able to obtain {\em necessary
and sufficient} conditions for classicality, the principal
tool we employ is the solution to the Stieltjes moment
problem~\cite{shohat-book} in the classical theory of
moments - when can a given sequence of moments arise from a
well-defined probability distribution?  The Stieltjes
problem deals with the case when the probability density is
defined over the half line $[0,\infty)\/$.  In our context,
this corresponds to the range of the intensity variable of
light.

The material of the paper is arranged as follows.
Section~II collects basic definitions related to the
diagonal coherent state representation of a single mode
field, and discusses the classical - nonclassical divide
from several angles.  The final one chosen is that suited
to phase insensitive measurements.  Both the probabilities
appearing in the PND, and the factorial moments of the PND,
are recognised as moments of two auxiliary distributions
over $[0,\infty)\/$, one with a clear physical meaning and
the other a formal one.  
In Section~III we give the basic method for
obtaining local classicality conditions on the PND and
obtain explicitly a minimal three term condition as an
example.  This condition involves the photon number
probabilities $p_n\/$ for three successive values of $n\/$, in
contradistinction to the condition in terms of the Mandel
$Q\/$ parameter which involves three successive (the zeroth
moment which equals unity in every state, and the 
first and second) moments of $p_n\/$.  The keen reader will
notice indications of an interesting duality already at
this initial stage.  We then put together a series of
remarks and preliminary results based on these minimal
conditions, which give a good orientation towards
understanding the features of the PND from both local and
global points of view.  In particular we are able to
discriminate between oscillations in the PND which
genuinely reflect nonclassicality and others which do not.

Section~IV  develops the
complete necessary and sufficient conditions for
classicality of the PND, by mapping the present problem on
to the Stieltjes moment problem which is very well known
and extensively studied.  All the conditions obtained here
are local in the sense mentioned above.  From the necessary
and sufficient conditions we then extract the conditions
for classicality which go one step beyond the minimal ones
discussed in Section~II; this helps us sharpen some of the
conclusions reached earlier.

Section~V gives an alternative or dual approach based on
the factorial moments of the PND.  This again is cast into
the form of a Stieltjes problem, and the necessary and
sufficient conditions for classicality are similar to the
previous ones in structure.  Section~VI compares and
connects the two treatments, and Section~VII contains
concluding remarks.

\section{Basic definitions, preliminary remarks and results}
We consider a single mode quantized radiation field with
photon creation and annihilation operators
$\hat{a}^{\dag},\hat{a}\/$ obeying the standard commutation
relation
\begin{eqnarray}
\protect[\hat{a},\hat{a}^{\dag}\protect] \equiv
\hat{a}\hat{a}^{\dag} -\hat{a}^{\dag}\hat{a} = 1.
\end{eqnarray} 
A general (pure or mixed) state of the field is described
by a hermitian nonnegative density operator $\hat{\rho}\/$
with unit trace.  According to the diagonal coherent state
representation theorem, $\hat{\rho}\/$ can be expanded as an
integral over projections on to the coherent states:
\begin{eqnarray}
\hat{\rho} = \int\frac{d^2 z}{\pi} \phi(z) |z\rangle \langle z|\;,
\end{eqnarray}
the integration being over the entire complex plane.  Here
the coherent state $|z\rangle \/$ is the normalized right eigenstate
of $\hat{a}\/$ with (generally complex) eigenvalue $z\/$,
related to the number operator eigenstates (Fock states)
$|n\rangle \/$ in the standard manner:
\begin{eqnarray}
|z\rangle  &=& \exp \left(-\frac{1}{2} |z|^2\right)
\sum\limits^{\infty}_{n=0}
\frac{\displaystyle z^n}{\displaystyle \sqrt{n!}}|n\rangle 
\nonumber\\
&=& \exp \left(-\frac{1}{2}|z|^2 +
z\hat{a}^{\dag}\right)|0\rangle \;,
\nonumber\\
|n\rangle &=& \left(\hat{a}^{\dag}\right)^n|0\rangle /\sqrt{n!}\;;
\nonumber\\
\hat{a}|z\rangle  &=& z|z\rangle \;,\;\hat{a}^{\dag}\hat{a}|n\rangle  = n|n\rangle \;.
\end{eqnarray}
The weight function $\phi(z)\/$ in eqn.(2.2) is real and
normalised to unit integral:
\begin{eqnarray}
\int\frac{d^2z}{\pi} \phi(z) =1\;.
\end{eqnarray}
However it is in general not pointwise
nonnegative~\cite{hillery-pla-1985}, and can be quite a
singular quantity, namely a member of a certain precisely
defined class of distributions over the plane.

Any (hermitian) observable can always be written as a
function of $\hat{a}^{\dag}\/$ and $\hat{a}\/$ in normal
ordered form, $F\left(\hat{a}^{\dag},\hat{a}\right)\/$
say~\cite{chaill-pr-1969+}.  Its expectation value in the
state $\hat{\rho}\/$ is then given by
\begin{eqnarray}
\langle F\left(\hat{a}^{\dag},\hat{a}\right)\rangle  &=&
\mbox{Tr}\left(\hat{\rho}
F\left(\hat{a}^{\dag},\hat{a}\right)\right)\nonumber\\ &=&
\int\frac{d^2 z}{\pi} \phi(z) F(z^*,z)\;.
\end{eqnarray}
If in particular $F\left(\hat{a}^{\dag},\hat{a}\right)\/$ is
phase invariant, then its expectation value does not
require all the ``information'' contained in $\phi(z)\/$, and
a simpler angle averaged auxiliary distribution ${\cal
P}(I)\/$ suffices:
\begin{eqnarray} 
F\left(\hat{a}^{\dag} e^{i\alpha},\hat{a}
e^{-i\alpha}\right)&=& F\left(\hat{a}^{\dag},\hat{a}\right)
\Rightarrow\nonumber\\
\langle F\left(\hat{a}^{\dag},\hat{a}\right)\rangle 
&=&\int\limits^{\infty}_{0} dI {\cal P}(I)
F\left(I^{1/2},I^{1/2}\right)\;,\nonumber\\ {\cal P}(I) &=&
\int\limits^{2\pi}_{0} \frac{d\theta}{2\pi} \phi
\left(I^{1/2} e^{i\theta}\right)\;,\nonumber\\
\int\limits^{\infty}_{0} dI{\cal P}(I) &=& 1.
\end{eqnarray}
We can regard ${\cal P}(I)\/$ as a real normalised marginal
radial distribution function obtained from the complete
$\phi(z)\/$. Note that $F(\hat{a}^{\dagger},\hat{a})\/$ being
phase invariant is the same as it being a function of the number
operator $\hat{a}^{\dagger}\hat{a}\/$.

An important set of phase invariant $F$'s leads to the
photon number distribution (PND) or photon number
probabilities $\{p_n\}\/$ in the state $\hat{\rho}$:
\begin{eqnarray}
F_n\left(\hat{a}^{\dag},\hat{a}\right) &=&
:\;e^{-\hat{a}^{\dag}\hat{a}}
\frac{\left(\hat{a}^{\dag}\hat{a}\right)^n}
{n!} \;:\; = |n\rangle \langle  n|\Rightarrow\nonumber\\ p_n &=&
\langle F_n\left(\hat{a}^{\dag},\hat{a}\right)\rangle \nonumber\\
&=&\langle n|\hat{\rho}|n\rangle \nonumber\\ &=&\int\limits^{\infty}_{0}
dI {\cal P}(I) e^{-I} I^n/n!\;\;\;, n = 0, 1, 2,\ldots\;.
\end{eqnarray}
(The colons denote normal ordering).  These $p_n\/$ are
always well-defined for any bonafide $\hat{\rho}\/$ and
always obey the laws for a discrete probability
distribution:
\begin{eqnarray} 
p_n \geq 0,\;\;\sum\limits^{\infty}_{n=0} p_n = 1.  
\end{eqnarray}

With the two quantities $\phi(z)\/$ and ${\cal P}(I)\/$ in
hand, we can set up a three-fold classification of states
$\hat{\rho}\/$ of steadily increasing nonclassicality.
Conventionally $\hat{\rho}\/$ is said to be classical if
$\phi(z)\/$ itself is a probability distribution, that is,
pointwise nonnegative and nowhere more singular than a
delta function; otherwise it is nonclassical.  In the
former case it follows that ${\cal P}(I)\/$ also can be
interpreted as a probability distribution for the
intensity.  In the latter case, a further refinement is
possible based on the properties of ${\cal P}(I)\/$ and we
arrive at the following scheme:
\begin{eqnarray}
\hat{\rho}\;\mbox{classical}&\Leftrightarrow& \phi(z), {\cal P}(I) 
\geq 0\; ;\nonumber\\ \hat{\rho}\;\mbox{weakly
nonclassical}&\Leftrightarrow& \phi(z)\not{\geq} 0\;,\;
{\cal P}(I)\geq 0\;;\nonumber\\ \hat{\rho}\;\mbox{strongly
nonclassical}&\Leftrightarrow& \phi(z) \not{\geq}
0\;,\;{\cal P}(I)\not{\geq} 0\;.
\end{eqnarray}
This is an exhaustive and mutually exclusive
classification.  The special role or significance of the
PND $\{p_n\}\/$ can now be expressed as follows: {\em even if
${\cal P}(I)\/$ is not a well-defined probability
distribution and we are in the strongly nonclassical
regime, the quantity ${\cal P}(I)e^{-I}\/$ always has
well-defined finite moments $n! p_n\/$ for all $n$}.  For
ease in the following we introduce a special symbol for
these moments,
\begin{eqnarray} 
q_n = n! p_n&=&
\int\limits^{\infty}_{0} dI
\tilde{{\cal P}}(I)\; I^n\;,\nonumber\\ 
\tilde{{\cal P}}(I)
&=& {\cal P}(I) e^{-I}\;,
\end{eqnarray}
the dependence on $\hat{\rho}\/$ being left implicit.  On the
other hand, the PND cannot discriminate between the
classical and the weakly nonclassical cases in eqn.(2.9),
in both of which ${\cal P}(I)\geq 0\/$.  Indeed, given a PND
$\{p_n\}\/$ leading to a pointwise nonnegative ${\cal P}(I)\/$
(see below), if no phase sensitive quantities are to be
measured we may take
\begin{eqnarray} 
\phi(z) = {\cal P}(|z|^2), 
\end{eqnarray}
and check that via eqn.(2.2) we get a physically acceptable
density operator $\hat{\rho}$.

An interesting illustration of the weakly nonclassical case
of eqn.(2.9) is the one-parameter family of states that
results when a coherent state $|z_0\rangle \/$ evolves through a
Kerr medium for a time interval $t\/$.  Indeed, for suitable
$t\/$ the state that results is the Yurke-Stoler
state~\cite{yurke-prl-1986}
\begin{eqnarray} 
|\psi\rangle  =
\frac{1}{\sqrt{2}} (|z_0\rangle  \pm i|-z_0 \rangle )\;.
\end{eqnarray}
This being the superposition of two coherent states has a
$\phi(z)\/$ more singular than a tempered distribution.
Nevertheless when angle averaged it leads to the same
${\cal P}(I)\/$ (and hence the same PND) as for the single
coherent state $|z_0\rangle \;:\;{\cal P}(I) =\delta(I-|z_0|^2)\/$.
The Hamiltonian of the Kerr medium being a function of
$\hat{a}^{\dag}\hat{a}\/$ leaves the diagonal matrix elements
$p_n = \langle n|\hat{\rho}|n\rangle \/$ unaffected, but changes only the
phases of $\langle m|\hat{\rho}|n\rangle \/$ for $m\neq n\/$, and ${\cal P}(I)\/$
depends only on the diagonal elements of $\hat{\rho}\/$ and not
on the off-diagonal elements.
  
In the remainder of this work we shall be exclusively
concerned with the PND $\{p_n\}\/$, or more generally with
expectation values of phase insensitive observables alone.
For these purposes we shall combinedly refer to the
classical and weakly nonclassical cases of eqn.(2.9) as
{\em classical} $({\cal P}(I)\geq 0)\/$; and to the strongly
nonclassical in eqn.(2.9) as {\em nonclassical} 
$({\cal P}(I) \not{\geq} 0)\/$. That is, for phase insensitive
observables, we define:
\begin{eqnarray}
\hat{\rho}\;\mbox{classical}&\Leftrightarrow&{\cal P}(I) \geq 0\;;
\nonumber\\
\hat{\rho}\;\mbox{nonclassical}&\Leftrightarrow& {\cal P}(I) 
\not{\geq} 0\;. 
\end{eqnarray}
Our principal aim now will be to develop {\em necessary and
sufficient} conditions on the PND $\{p_n\}\/$ for
$\hat{\rho}\/$ to be classical; we may then say for brevity
that we have a classical PND.

The recent experiment of Munroe et al~\cite{munroe} best illustrates
our considerations based on phase-insensitive
measurements on the one hand and our classical-nonclassical 
divide~(2.13) based on ${\cal P}(I)\/$ on the other. They show
that the phase averaged quadrature amplitude distribution 
$\overline{P}(\xi)\/$ measured by their optical homodyne 
detection apparatus is invertibly related to the PND.
We shall show presently that the relationship~(2.7) between 
${\cal P}(I)\/$ and the PND is invertible. It thus follows that
their probability $\overline{P}(\xi)\/$ and our quasiprobability 
${\cal P}(I)\/$ are invertibly related: their experiment extracts 
no more, and no less, information than contained in ${\cal P}(I)\/$.
In other words, the classification~(2.13), rather than~(2.9), is the
one relevant for such experiments.

The distribution character of $\phi(z)\/$ passes over to a
corresponding property for ${\cal P}(I)\/$ - namely, it too
is a member of a precisely defined class of distributions
over $[0,\infty)\/$.  While it is clear that the sequence
$\{p_n\}\/$ cannot possibly capture all the information
contained in $\phi(z)\/$ as all the off-diagonal matrix
elements $\langle m|\hat{\rho}|n\rangle \;,\;m\neq n\/$, are ignored, we
can easily show that the PND $\{p_n\}\/$ and the distribution
${\cal P}(I)\/$ determine each other uniquely.  To recover
${\cal P}(I)\/$ from $\{p_n\}\/$ we define a generating
function $\Lambda(K),\;0\leq K<\infty\/$, to represent the
latter:
\begin{eqnarray} 
\Lambda(K) &=&
\sum\limits^{\infty}_{n=0} (-K)^n p_n /n!\nonumber\\
&=& \int\limits^{\infty}_{0} dI {\cal P}(I)\; e^{-I}
J_0(2\sqrt{IK})\;.
\end{eqnarray}
There is a great deal of freedom in the way we set up
$\Lambda(K)\/$; we have chosen it so that, among other
things, on the basis of eqn.(2.8) it is entire analytic in
$K\/$.  One can then invert eqn.(2.14) by using the Fourier -
Bessel integral theorem to get
\begin{eqnarray}
 {\cal P}(I)
= e^{I}\int\limits^{\infty}_{0} dK\;\Lambda(K)
J_0(2\sqrt{KI})\;.
\end{eqnarray}

Several signatures of nonclassicality of the PND are well
known.  The most familiar is the Mandel Q-parameter
criterion which distinguishes (in a global sense) between
super and subpoissonian PND's:
\begin{eqnarray*} Q &=&
((\Delta n)^2 -\langle n\rangle )/\langle n\rangle \\ &=& (\langle n^2\rangle  - \langle n\rangle ^2 -
\langle n\rangle )/\langle n\rangle \;,\\ \langle n\rangle  &=&\sum\limits^{\infty}_{n=0} n\;p_n =
\int\limits^{\infty}_{0} 
dI{\cal P}(I) I\;,\\ \langle n^2\rangle  &=& \sum\limits^{\infty}_{n=0}
n^2 p_n = \int\limits^{\infty}_{0} dI {\cal P}(I) I(I+1)\;;
\end{eqnarray*}

\begin{eqnarray}
&\mbox{classical PND} \Rightarrow Q\geq
0\;,\;\mbox{superpoissonian statistics}\;,\nonumber&\\ 
&Q < 0 \Rightarrow  \mbox{nonclassical PND,subpoissonian
statistics}&.\nonumber
\end{eqnarray}
\vspace*{-24pt}
\begin{equation}
\end{equation}
However one can see that there are (uncountably many)
states $\hat{\rho}\/$ for which $Q\/$ is undefined, on account
of either $\langle n^2\rangle \/$ or both $\langle n\rangle \/$ 
and $\langle n^2\rangle \/$ being
divergent. This can happen because $Q\/$ involves $p_n\/$ for
all $n\/$, and the rate of decrease of $p_n\/$ as $n\rightarrow
\infty\/$ may be insufficient
for convergence of either $\langle n\rangle \/$ or $\langle n^2\rangle $.  A similar
situation obtains with the factorial moments of the PND,
namely
\begin{eqnarray} 
\gamma_m&=& \langle \hat{a}^{\dag m}\hat{a}^m \rangle 
\nonumber\\ &=& \sum\limits^{\infty}_{n=m}
\frac{n!}{(n-m)!} p_n\nonumber\\
&=& \int\limits^{\infty}_{0} dI {\cal P}(I)\; I^m\;,\; \;
m=0,1,,2,\ldots\;.  
\end{eqnarray}
In terms of these, the
Mandel parameter is
\begin{eqnarray} 
Q = \left(\gamma_2 -\gamma^2_1\right)/\gamma_1\;.  
\end{eqnarray}
A well-known feature of the factorial moments is the
following~\cite{klauder-book}:
\begin{eqnarray} 
\mbox{Classical PND}&\Rightarrow&
\gamma_{m^{\prime}} \gamma_m \leq
\gamma_{m^{\prime}+m}\leq 
\left(\gamma_{2m^{\prime}}\gamma_{2m}\right)^{1/2}\;,\nonumber\\
&& m^{\prime}, m=0,1,2,\ldots\; .
\end{eqnarray}

However, as with $Q,\;\gamma_m\/$ also involves $p_n\/$ for all
but a finite number of values of $n\/$, and is undefined for
the vast majority of states $\hat{\rho}$.

This discussion shows that in general, in contrast to
$\tilde{{\cal P}}(I)={\cal P}(I)\;e^{-I}\/$, the moments
$\gamma_m\/$ of ${\cal P}(I)\/$ may not all exist.  States
$\hat{\rho}\/$ for which all $\gamma_m\/$, or even all
factorial moments upto some fixed  maximum order, are finite are
naturally quite severely restricted.

One may enquire about the properties of the ordinary
moments of the PND, namely 
\begin{eqnarray}
 \delta_m&=&
\langle \left(\hat{a}^{\dag}\hat{a}\right)^m\rangle \nonumber\\ &=&
\sum\limits^{\infty}_{n=0} n^m p_n\nonumber\\
&=& \int\limits^{\infty}_{0} dI{\cal P}(I)\; e^{-I}
\sum\limits^{\infty}_{n=0} n^m I^n/n!\nonumber\\
&=&\int\limits^{\infty}_{0} dI\; {\cal P}(I)\; e^{-I}\;
\left(I\;\frac{d}{dI}\right)^m\;e^{I}\;,\;\nonumber \\
&m&=0,1,2,\ldots .
\end{eqnarray}
While they may seem to behave qualitatively like the
$\gamma_m\/$, they are however not expressible as the moments
of some function of $I\/$ simply related to ${\cal P}(I)\/$ (as
the $p_n\/$ and the $\gamma_m\/$ are, eqns.(2.7,10,17)).
Therefore we do not deal with them hereafter.

\section{Local Classicality Conditions on 
$\protect\lowercase{\large p}_{\protect\lowercase{n}}\/$}
\setcounter{equation}{0}
The above discussion motivates the search for {\em local}
consequences of classicality on the PND, that is,
inequalities involving only a finite number of the $p_n\/$'s
which are always well defined and which must be obeyed if
the PND is classical.  Based on eqn.(2.7), the general form
of such conditions can be surveyed as follows.  Consider a
finite degree polynomial $f(x)\/$ in a real nonnegative
variable $x\/$, with the property of itself being real
nonnegative:
\begin{eqnarray} 
f(x) =\sum\limits^{N}_{n=0} c_n\;x^n\geq 0,\quad\mbox{for}\;
0\leq x < \infty\; .
\end{eqnarray}
The nontrivial case here is when some coefficients $c_n\/$
are negative; for example $f(x)\/$ could be the square of a
polynomial with coefficients of both signs.  Then we find 
in view of~(2.10):
\begin{eqnarray}
{\cal P}(I)\geq 0 \Rightarrow \int\limits^{\infty}_{0} dI
{\cal P}(I) e^{-I} f(I) = \sum\limits^{N}_{n=0}
c_n\;q_n\geq 0\; .
\end{eqnarray}
One thus finds that for a classical PND, every polynomial
obeying eqn.(3.1) leads to one inequality (3.2); so these
are necessary local conditions for classicality.

One can easily check that there are no such local
conditions involving only two consecutive $p_n$'s.  The
simplest or minimal local conditions involve three
consecutive $p_n$'s, and are obtained by choosing for
$f(x)\/$ an integral power of $x\/$ times a quadratic in the
form of a perfect square:
\begin{eqnarray}
{\cal P}(I) \geq 0\;,\;f(x)= x^{n-1}(1-ax)^2,
a\;\mbox{real}\;
\Rightarrow\nonumber\\
q_{n-1} - 2a q_n + a^2 q_{n+1} \geq 0\;,\;\mbox{all
real}\;a \Rightarrow \nonumber\\ q_n^2 \leq q_{n-1}
q_{n+1}\;,\; n=1,2,\ldots .
\end{eqnarray}
Written in term of the true probabilities $p_n\/$ this reads:
\begin{eqnarray}
\mbox{PND classical}&\Rightarrow& p^2_n \leq \left(1
+\frac{1}{n}\right) p_{n-1} p_{n+1},\nonumber \\ 
&& n=1,2,\ldots .
\end{eqnarray}
We shall call this {\em sequence of local conditions the minimal
or three-term classicality conditions} on the PND.

We now list a series of direct consequences of the above
discussions and of the minimal local conditions (3.3) for
a PND to be classical.
\begin{itemize}
\item[(a)] For a classical PND, the (nonnegative) 
distribution ${\cal P}(I)\/$ is
either concentrated at $I=0,\;{\cal P}(I) =\delta(I)\/$, or
else it has positive definite weight for some values or
ranges of $I\/$ strictly greater than zero.  The former
corresponds to the vacuum state:
\begin{eqnarray}
 {\cal P}(I) =\delta(I) &\Rightarrow& p_n
=q_n=\delta_{n,0}\;,\nonumber\\ \hat{\rho} &=& |0\rangle \langle  0|\; .
\end{eqnarray}
In the latter situation we have
\begin{eqnarray}
 {\cal P}(I) 
\geq 0,\;{\cal P}(I) &\neq& \delta(I) \Rightarrow p_n > 0,\;
\nonumber \\
n&=&0,1,2,\ldots .  
\end{eqnarray}
Both~(3.5) and~(3.6) are immediate consequences of~(2.7)
or, equivalently,~(2.10).
 Therefore we conclude:
(i) if $p_0=0\/$, the PND is definitely nonclassical; (ii) if
the PND is classical and the state is not the vacuum, every
$p_n\/$ is strictly positive; equally well a non-vacuum
classical state cannot be orthogonal to any Fock state;
(iii) conversely,for a nonvacuum state, the vanishing of
any one (or more) of the $p_n$'s implies nonclassicality.

These conclusions can also be obtained recursively from the
minimal conditions (3.3): if $q_{n_{0}}=0\/$ for some
$n_0\geq 1\/$, repeated use of these inequalities leads to
$q_n=0\/$ for all $n\geq 1\/$, the state being assumed to be
classical.

Hereafter we omit the vacuum state from consideration.

\item[(b)] As a corollary to the above, we see that if $\hat{\rho}\/$
is any state and $\hat{\rho}^{\prime}\/$ is obtained by
adding some number $m\/$ of photons to $\hat{\rho}\/$,
\begin{eqnarray} 
\hat{\rho}^{\prime} = {\cal N}^{-1}
\hat{a}^{\dag m}\hat{\rho} 
\hat{a}^m\;,\; m\geq 1\;, 
\end{eqnarray}
then $\hat{\rho}^{\prime}\/$ is always nonclassical.
The normalisation constant ${\cal N}\/$ needs to be 
finite for~(3.7) to make sense as a state.  A direct
calculation gives:
\begin{eqnarray} 
{\cal N} &=& \mbox{tr}(\hat{a}^{m}\hat{a}^{\dagger}{}^{m}
\hat{\rho})=\sum\limits^{\infty}_{n=0}
\frac{(n+m)!}{n!} p_n\;;\nonumber\\
p^{\prime}_n &=& \left\{\begin{array}{ll}0,&n\leq
m-1\;,\\ {\cal N}^{-1} \frac{n!}{(n-m)!} p_{n-m},& n\geq
m\;, \end{array}\right.
\end{eqnarray}
where $\hat{\rho}\/$ determines the PND $\{p_n\}\/$ and
$\hat{\rho}^{\prime}\/$ determines $\{p^{\prime}_{n}\}\/$.  If
the series for ${\cal N}\/$ converges, then
$\hat{\rho}^{\prime}\/$ is a well defined state; and then the
vanishing of $p^{\prime}_0, p^{\prime}_1,\ldots,
p^{\prime}_{m-1}\/$ establishes its nonclassicality.  This
proves that {\em all photon added states are nonclassical}.
Nonclassicality of photon added coherent
states~\cite{agarwal-pra-1991} and photon added thermal
states~\cite{agarwal-pra-1992,jones-quclopt-1997} (${\cal
N}\/$ is finite in both cases) have already been studied in
great detail

\item[(c)] For any classical PND, since $q_0=p_0 > 0\/$, we see that
$q_0^{-1} \tilde{\cal P}(I)\/$ is a nonnegative function
normalised to unit integral. Therefore it can be treated
mathematically as though it were a probability distribution
over $[0,\infty)\/$.  However its physical interpretation is
quite different from that of ${\cal P}(I)\/$ given earlier.

\item[(d)] Since $\{q_n\}\/$ is a geometric sequence for a Poissonian
distribution, we see that the minimal classicality
conditions (3.3) are saturated in this case for every $n\/$.
Thus we can interpret these conditions as requiring that
the sequence $\{p_n\}\/$ be {\em locally} Poissonian or
superpoissonian {\em at each $n\/$}.  For a classical PND
each $q_n\/$ (respectively $p_n\/$) has a lower bound
determined by the values of the two previous $q\/$'s
(respectively $p\/$'s):
\begin{eqnarray} 
q_n\geq q^2_{n-1} /q_{n-2}\;,\; p_n&\geq&
\left(1-\frac{1}{n}\right) p^2_{n-1}
/ p_{n-2}\;,\;
\nonumber \\
 n&\geq 2&\; .
\end{eqnarray}
If even at one value of $n\geq 1\/$ we have
 \begin{eqnarray}
q^2_n > q_{n-1} q_{n+1}\;, 
\end{eqnarray} 
so that the PND
is locally subpoissonian at that value of $n\/$, the state is
definitely nonclassical.

\item[(e)] The inequalities (3.3) also imply that for a classical
PND we can never have $q_n > q_{n+1}\;,\;q_{n-1}\/$ for any
$n \geq 1\/$.  That is, for classical states {\em local
maxima in the sequence $\{q_n\}\/$ are ruled out}.  This
implies in particular that no oscillations in $\{q_n\}\/$ are
possible for such states.  Thus we arrive at the important
conclusion: {\em oscillation in} $\{q_n\}\/$ {\em is a sure
sign of nonclassicality}.

\item[(f)] On the other hand, the minimal conditions (3.3) permit
a local minimum in $\{q_n\}\/$ but no more than one.  This is
because if we have minima at $n_1\/$ and again at $n_2\geq
n_1+2\/$, there would necessarily be a local maximum in
between, which is disallowed for classical states.
\vspace*{-0.5cm}
\begin{figure}
\hspace{0.25cm}\epsfxsize=7cm
\epsfbox{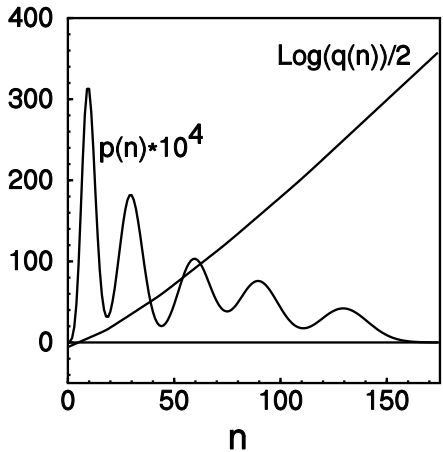}
\caption{``Classical'' oscillation in the photon number distribution,
for an incoherent mixture of coherent states with density
matrix $\rho =\/$ $\lambda_1 \vert \alpha_1 \rangle \langle
\alpha_1 \vert+ \/\/$
$\lambda_2 \vert \alpha_2 \rangle \langle \alpha_2 \vert+
\/\/$
$\lambda_3 \vert \alpha_3 \rangle \langle \alpha_3 \vert+
\/$
$\lambda_4 \vert \alpha_4 \rangle \langle \alpha_4
\vert+\/$
$\lambda_5 \vert \alpha_5 \rangle \langle \alpha_5 \vert\/$
with $|\alpha_1|^2=10\/$, $|\alpha_2|^2=30\/$,
$|\alpha_3|^2=60\/$, $|\alpha_4|^2=90\/$, $|\alpha_5|^2=130\/$
and $\lambda_1=\lambda_2=0.25\/$, $\lambda_3=0.2\/$,
$\lambda_4=0.18\/$ $\lambda_5= 0.12\/$. The symbols $p(n)$, $q(n)\/$
stand, respectively, for $p_n, q_n\/$ of the text. It may be noted
that $q_n\/$ exhibits no oscillations for this classical state.}
\end{figure}

\item[(g)] We can now combine (e) and (f) above to state the
following.  Ever since the important work of Schleich and
Wheeler on interference in phase
space~\cite{schleich-nature-1987+}, the statement that
oscillation in $\{p_n\}\/$ is a signature of nonclassicality
has almost become a folklore.  Indeed, oscillations in
$\{p_n\}\/$ are known as nonclassical
oscillations~\cite{caves-pra-1991}.  The striking virtue of
this characterization is that it is {\em local} in $n\/$, in
the spirit of our present approach, as against other
characterizations based on the (factorial) moments of
$\{p_n\}\/$.  We have shown in Fig.1 the sequence $\{p_n\}\/$
for a suitably chosen incoherent superposition of coherent
states.  This state is classical by construction, yet it
exhibits oscillations in $\{p_n\}\/$, showing that the above
characterization needs quantification of some sort while
retaining the attractive feature of being local in $n\/$.
Our minimal local conditions (3.3) can be viewed as a
quantification of this type.  They limit the extent to
which oscillations can occur in $\{p_n\}\/$ if the state is
classical.  These limits are obtained by translating
properties (e), (f) of $\{q_n\}\/$ above into corresponding
properties of $\{p_n\}\/$.  For classical $\{q_n\}\/$, (e) and
(f) allow only four generic patterns of behaviour: (i)
$q_n\/$ nondecreasing, provided a constant phase (if any) is
{\em followed}, {\em not preceded}, by an increasing phase
(if any); (ii) $q_n\/$ non increasing, provided a constant
phase (if any) is {\em preceded, not followed}, by a
decreasing phase (if any); (iii) $q_n\/$ constant (actually
subsumed in the two previous possibilities); (iv) $q_n\/$
passing through one local minimum, monotonic decreasing
(increasing) before (after) the minimum.  The monotonic
increasing case is part of (i), the monotonic decreasing
case is part of (ii).  Well known classical states
constitute examples of these generic behaviours of
classical $\{q_n\}\/$, as shown in Fig.2.
\end{itemize}
\vspace*{-0.5cm}
\begin{figure}
\hspace{0.25cm}\epsfxsize=7cm
\epsfbox{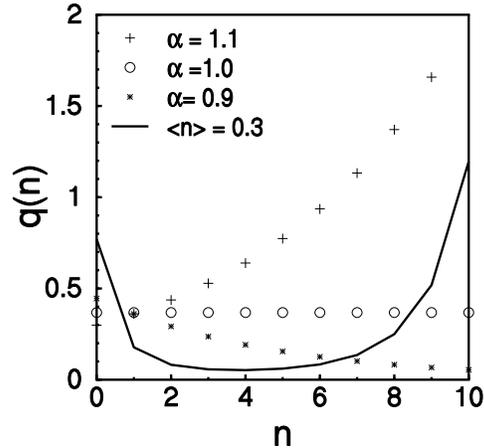}
\caption{Generic behaviours of the sequence 
$\{q_n\}\/$ for different classical
states. First three are coherent states with different
displacements and the fourth one is a thermal state.}
\end{figure}

It turns out that constant phases in $\{q_n\}\/$ in cases (i)
and (ii) above, allowed by the minimal local conditions
(3.3), but not shown in Fig.2, are indeed forbidden by the
higher order local conditions as will be shown in
Section~IV.

All these restrictions on classical $\{q_n\}\/$ translate
into allowed behaviours for classical $\{p_n\}\/$.  First,
local maxima, $p_n>p_{n\pm 1}\/$, are permitted, provided
\begin{eqnarray}
\frac{\displaystyle p_n}{\displaystyle p_{n-1}}\;
\frac{\displaystyle p_n}{\displaystyle p_{n+1}}
\leq 1+\frac{1}{n}\;.
\end{eqnarray}
Next, again because of the factor
$\left(1+\frac{1}{n}\right)\/$ on the right hand side in the
minimal classicality condition (3.4), some amount of
oscillation in $\{p_n\}\/$ is allowed.  This is just the kind
of oscillation seen in Fig.1.  Note that the period of
oscillation (difference of $n\/$ values at two successive
maxima in $\{p_n\}\/$) in Fig.1 is substantially greater than
two as against the period two oscillations occurring in the
PND of a squeezed or cat state. For period two oscillations
the conditions (3.4) place substantial restrictions on the
amplitude.  (Indeed the next higher order local conditions
to be derived in Section~IV make these restrictions even
more stringent).  Further, as the factor
$\left(1+\frac{1}{n}\right)\/$ approaches unity as $n\/$
becomes large, the restrictions on the amplitude of a
classically allowed period two oscillation are stronger at
higher values of $n\/$.  To end the discussion in this
paragraph we note that the {\em period two classical oscillation}
in $\{p_n\}\/$, allowed by our lowest order local condition
(3.3,4), indeed survives all higher order local
conditions, as can be seen by the example shown in Fig.3.
\vspace*{-0.5cm}
\begin{figure}
\hspace{0.25cm}\epsfxsize=7cm
\epsfbox{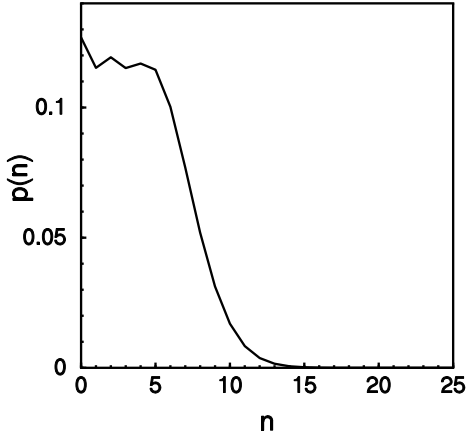}
\caption{Limits to PND oscillations from classicality conditions-
an example of incoherent mixture of three coherent states: $\rho =
\lambda_1 \vert \alpha_1 \rangle \langle \alpha_1 \vert
+\lambda_2 \vert \alpha_2 \rangle \langle \alpha_2 \vert
+\lambda_3 \vert \alpha_3 \rangle \langle \alpha_3
\vert\/$
with $|\alpha_1|^2=0.0\/$, $|\alpha_2|^2=1.56\/$, $|\alpha_3|^2=5.42\/$, 
$\lambda_1=0.06\/$, $\lambda_2=0.305$ and $\lambda_3=0.635\/$.
Period two oscillations though restricted 
in amplitude can  be present for a classical state.}
\end{figure}

An instructive example illustrating many of the points made
above is provided by the case of a pure state obtained as
the superposition of two ``opposite'' coherent states:
\begin{eqnarray}
|\psi(z_0,\theta)\rangle  &=& {\cal N}^{-1/2} \left( |z_0\rangle 
+ e^{i\theta}| -z_0\rangle  \right)\;,\nonumber\\ {\cal N}
&=& 2\left(1 + e^{-2|z_{0}|^{2}} \cos \theta\right).
\end{eqnarray}
Here $\theta\/$ is the relative phase (in the Pancharatnam
sense~\cite{panchratnam-indian-acad-1956+}) between the two
components of the superposition.  As special cases we get
the Yurke-Stoler states (2.12) when $\theta=\pm
\pi/2\/$, and the cat states when $\theta=0,\pi\/$.  For the PND we
have
\begin{eqnarray} 
p_n&=&|\langle n|\psi(z_0,\theta)\rangle |^2\nonumber\\ &=&
e^{-|z_{0}|^{2}}\; \frac{\displaystyle
|z_{0}|^{2n}}{\displaystyle n!}\; \frac{\displaystyle
(1+(-1)^n\cos\theta)} {\displaystyle
\left(1+e^{-2|z_{0}|^{2}}\cos
\theta\right)}\;,\; 
\end{eqnarray}
so that
\begin{eqnarray} 
q_{n-1} q_{n+1}/q^2_n &=&
\left\{(1-(-1)^n\cos\theta)/(1+(-1)^n\cos\theta)
\right\}^2\nonumber\\ &\equiv& f_n(\theta)\;.
\end{eqnarray}
Clearly, $f_n(\theta)< 1\/$ for even $n\/$ if $-\pi/2 <\theta <
\pi/2\/$; and $f_n(\theta)< 1\/$ for odd $n\/$ if $\pi/2 <\theta
<3\pi/2\/$.  Thus the state $|\psi(z_0,\theta)\rangle \/$ violates the
minimal local condition (3.3) and is nonclassical for all
$\theta\neq \pm\pi/2\/$.  For $\theta=\pm\pi/2\/$ these
conditions are saturated at every $n\/$, and we have a
Poissonian $\{p_n\}\/$.

It is clear that nonclassicality of the Yurke-Stoler state
(2.12) can be exhibited only through phase sensitive
considerations, since the PND is poissonian. It is also
well known~\cite{hillery-pla-1985} that all pure states
other than coherent states are nonclassical at least at the
phase sensitive level.  What is really interesting is the
fact that for every $\theta\neq\pm\pi/2\/$, the
nonclassicality of $|\psi(z_0,\theta)\rangle \/$ is coded in the
phase insensitive PND, and that our minimal conditions
(3.3,4) capture this nonclassicality!

We conclude this Section with one more application of the
local condition.  Recent years have witnessed a remarkable
progress in quantum state reconstruction using techniques
of optical homodyne tomography~\cite{vogel-pra-1989+}. As
a consequence it is now possible to `map out' the Wigner
distribution of a state using the inverse Radon transform,
or reconstruct the density matrix in the Fock basis using a
set of pattern functions. Schiller et
al~\cite{schiller-prl-1996} report such a reconstructed
density matrix $\rho_{m,n}\/\/$ for $m,n
\leq 6\/\/$. The reported values of
$q_n\/$ for $n=0\/\/$ to $6\/$ are $0.44\/$, $0.07\/$,
$0.26\/$, $0.30\/$, $1.44\/$, $3.60\/$ and $28.80\/$. The
local conditions are clearly violated: $q_1\/q_3 <
q_2^2,\;q_3\/q_5 < q_4^2\/$. Hence the state is
nonclassical. The state was, of course,
 known to be quadrature squeezed,
and hence nonclassical. What is interesting about the
present illustration is the fact that the nonclassicality
of the Schiller et al state survives phase averaging, and
that we are able to arrive at this definitive conclusion
with just a few values of the diagonal elements of the
density matrix!
\section{Necessary and sufficient conditions for classicality of
the PND - the Stieltjes moment problem}
\setcounter{equation}{0}
We have seen in the previous Section that pointwise
nonnegativity of ${\cal P}(I)\/$, hence of $\tilde{\cal
P}(I)\/$, leads to the minimal local conditions (3.3,4) on
the sequences $\{q_n\}, \{p_n\}\/$. As already remarked,
these are necessary conditions for classicality.  Central
to this derivation was appreciation of the fact that the
$q_n\/$ are {\em moments} of $\tilde{\cal P}(I)\/$,
eqn.(2.10).  We have also seen that if the state is
classical, then $q_0^{-1}\tilde{\cal P}(I)\/$ is
well-defined and mathematically interpretable as a
probability distribution over $[0,\infty)\/$.  In this
Section we exhibit the {\em necessary and sufficient}
conditions on the PND $\{q_n\}\/$ in order that the state
$\hat{\rho}\/$ be classical.  We then use these conditions
to examine the extension of the inequalities (2..23,24) to
the five-term case.

The reconstruction of a probability distribution from its
moment sequence constitutes the classical moment problem on
which there exists an enormous amount of
literature~\cite{shohat-book}.  When the probability
distribution is over $[0,\infty)\/$, one calls it the
Stieltjes moment problem.  The Hamburger moment problem
corresponds to the case where the probability distribution
is over $(-\infty,\infty)\/$; and the Hausdorff or the {\em
little} moment problem corresponds to the finite interval
$[0,1]\/$.  Since the argument of $\tilde{\cal P}(I)\/$ is
nonnegative, our problem of deriving necessary and
sufficient conditions on the moment sequence $\{q_n\}\/$ in
order that $q_0^{-1}\tilde{\cal P}(I)\/$ is a true
probability distribution is then a Stieltjes moment
problem.  In this Section we can take advantage of the fact
that these moments $\{q_n\}\/$ of $\tilde{\cal P}(I)\/$ are
always well-defined and finite, whatever the state
$\hat{\rho}\/$ may be.

The solution of this classical problem is well known.  To
exhibit it, construct from the sequence $\{q_n\}\/$ two
symmetric matrices $L^{(N)},\tilde{L}^{(N)}\/$ of dimension
$(N+1)\/$ defined by
\begin{eqnarray} 
L_{mn}^{(N)} &=& q_{m+n}\;,\nonumber\\ \tilde{L}_{mn}^{(N)}
&=& q_{m+n+1}\;,
\;m,n=0,1,2,\ldots,\;N.
\end{eqnarray}
That is,
\begin{eqnarray} 
L^{(N)} &=&\left(\begin{array}{ccccc}
q_0&q_1&q_2&\ldots&q_N\\ q_1&q_2&q_3&\ldots&q_{N+1}\\
\vdots\\
q_N&q_{N+1}&q_{N+2}&\ldots&q_{2N}
\end{array}\right)\;,\nonumber\\
\tilde{L}^{(N)}&=&\left(\begin{array}{ccccc}
q_1&q_2&q_3&\ldots&q_{N+1}\\ q_2&q_3&q_4&\ldots&q_{N+2}\\
\vdots\\
q_{N+1}&q_{N+2}&q_{N+3}&\ldots&q_{2N+1} \end{array}\right).
\end{eqnarray}
Thus, $\tilde{L}^{(N)}\/$ arises from $L^{(N+1)}\/$ by deletion
of the first column and the last row in the latter.  Notice
also that the diagonal elements of $L^{(N)}\/$ are $q_n\/$ for
even $n\leq 2N\/$, while those of $\tilde{L}^{(N)}\/$ are $q_n\/$
for odd $n\leq 2N+1\/$.

Now we quote the fundamental classical
theorem~\cite{shohat-book}.

\noindent{\em \bf Theorem 1:}

The necessary and sufficient condition on the PND sequence
$\{q_n\}\/$ in order that the associated distribution
$q_0^{-1}\tilde{\cal P}(I)\/$ be a true probability
distribution over $[0,\infty)\/$ is that the above matrices
be nonnegative for all $N\geq 0\/$:
\begin{eqnarray}
\mbox{Classical PND}&\Leftrightarrow& {\cal P}(I)
\geq 0\Leftrightarrow
\tilde{\cal P}(I)\geq 0\nonumber\\ &\Leftrightarrow&
L^{(N)}, \tilde{L}^{(N)}\geq 0,\;N=0,1,2,\ldots .
\end{eqnarray}

\noindent{\em \bf Proof:}

We make use of the following two facts.

(i) A distribution $\tilde{\cal P}(I)\/$ over $[0,\infty)\/$ is
pointwise nonnegative if and only if it leads to a
nonnegative expectation value for every polynomial $f(I)\/$
which is itself pointwise
nonnegative over $[0,\infty)\/$~\cite{haviland-ajm-1934+}:
\begin{eqnarray}
&\tilde{\cal P}(I)\geq 0\;\mbox{for all}\; I\Leftrightarrow
\langle f(I)\rangle _{\tilde{\cal P}} = \int\limits^{\infty}_{0} dI
\tilde{\cal P}(I) f(I) \geq 0,&
\nonumber \\
&\mbox{for all polynomial}\;
f(I)\geq 0. &\!\!\!\!\!\!\!
\end{eqnarray}

(ii) Any polynomial $f(I)\/$ pointwise nonnegative over
$[0,\infty)\/$ can be written in terms of two perfect square
polynomials as~\cite{shohat-book}
\begin{eqnarray} 
f(I) = (f_1(I))^2 + I(f_2(I))^2\; .  
\end{eqnarray}

To prove necessity, suppose that $\tilde{\cal P}(I)\/$ is
pointwise nonnegative.  Consider the polynomial
$f_1(I)=\sum\limits^{N}_{n=0} c_n I^n\/$, where the $c_n\/$ are
arbitrary real coefficients.  By (4.4) we have
$\langle (f_1(I))^2\rangle _{\tilde{\cal P}}\geq 0\/$.  That is,
\begin{eqnarray}
\langle (f_1(I))^2\rangle _{\tilde{\cal P}} &=& \sum\limits^{N}_{m,n=0}
c_m c_n \langle  I^{m+n}\rangle _{\tilde{\cal P}} \nonumber\\
&=&\sum\limits^{N}_{m,n=0} c_m c_n q_{m+n}\nonumber\\
&=&\sum\limits^{N}_{m,n=0} c_m c_n L^{(N)}_{mn} \geq 0\; .
\end{eqnarray}
This means $L^{(N)}\geq 0\/$ for every $N\geq 0\/$.  Similarly,
writing $f_2(I)=\sum\limits^{N}_{n=0} d_n I^n\/$ and
evaluating the expectation value of the nonnegative
polynomial $I(f_2(I))^2\/$ we have from (4.4):
\begin{eqnarray}
\langle I(f_2(I))^2\rangle _{\tilde{\cal P}} &=& \sum\limits^{N}_{m,n=0}
d_m d_n \langle  I^{m+n+1}\rangle _{\tilde{\cal P}} \nonumber\\
&=&\sum\limits^{N}_{m,n=0} d_m d_n q_{m+n+1}\nonumber\\
&=&\sum\limits^{N}_{m,n=0} d_m d_n \tilde{L}^{(N)}_{mn}
\geq 0\; .
\end{eqnarray}
 This means $\tilde{L}^{(N)}\geq 0\/$ for every
$N\geq 0\/$.  Thus nonnegativity of $\tilde{\cal P}(I)\/$
implies that of every $L^{(N)}, \tilde{L}^{(N)}\/$.

To prove sufficiency, assume $L^{(N)},\tilde{L}^{(N)}\geq
0\/$ for $N\geq 0\/$.  Given any nonnegative polynomial $f(I)\/$,
writing it in the form (4.5) we find that
$\langle f(I)\rangle _{\tilde{\cal P}}\geq 0\/$.  This implies, by fact (i)
stated at the beginning of the proof, that $\tilde{\cal
P}(I)\geq 0\/$.  This completes the proof of the Theorem.

Now each matrix $L^{(N)}\/$ (respectively $\tilde{L}^{(N)}\/$)
arises from $L^{(N-1)}\/$ (respectively $\tilde{L}^{(N-1)}\/$)
by addition of one extra row and column, leaving the rest
intact.  Therefore the result (4.3) translates into a
series of determinant conditions:
\begin{eqnarray} 
D_N &=&\det L^{(N)}\;,\; \tilde{D}_N =\det
\tilde{L}^{(N)}\;:\nonumber\\
\mbox{Classical PND} &\Leftrightarrow& {\cal P}(I),
\tilde{\cal P}(I)
\geq 0\nonumber\\ &\Leftrightarrow& D_N\;,\;\tilde{D}_N
\geq 0\;,\; N=0,1,2,\ldots .
\end{eqnarray}

Assume now that we have a classical PND.  Then either $D_N,
\tilde{D}_N >  0\/$ for all $N\/$, or $D_N, \tilde{D}_N > 0\/$ for
$N\leq k\/$ and $D_N=\tilde{D}_N=0\/$ for $N>k\/$.  In the latter
case the support of $\tilde{\cal P}(I)\/$ (the set of values
of $I\/$ at which $\tilde{\cal P}(I)>0\/$) is a finite set of
$k\/$ points~\cite{widder-book}.  In the former case it is an
infinite set.  This can be intuitively understood along the
following lines.  Suppose $\tilde{\cal
P}(I)=\delta(I-I_0)\/$.  Then $q_n=I^n_0\/$ so that
$L_{mn}^{(N)}=I^m_0 I^n_0\/$ and $\tilde{L}^{(N)}_{mn}=I_0
I^m_0 I^n_0\/$.  That is, $L^{(N)}\/$ and $\tilde{L}^{(N)}\/$ are
(essentially) projection matrices.  Thus when the support
of $\tilde{\cal P}(I)\/$ is a finite set of points, $L^{(N)}\/$
(as also $\tilde{L}^{(N)}\/$) is the sum of $k\/$ projections.
Since $I_0=0\/$ contributes a projection only to $L^{(N)}\/$
but not to $\tilde{L}^{(N)}\/$, one has the following
refinement: if the support of $\tilde{\cal P}(I)\/$ consists
of $k\/$ points including the point $I=0\/$, then $D_N>0
(\tilde{D}_N>0)\/$ if and only if $N\leq k(N\leq k-1)\/$.

It is useful to remark that the first fact (4.4) we have
used in the proof of the theorem is common for all the
three types of moment problems.  What changes from one
moment problem to another is the second fact dealing with
the decomposition of nonnegative polynomials into sums of
square polynomials, thus enabling us to convert (4.4) into
simple matrix conditions.  For instance, for the Hamburger
moment problem on $(-\infty,\infty)\/$ we have, in place of
(4.5), the statement that every polynomial nonnegative over
$(-\infty,\infty)\/$ can be written as the sum of two square
polynomials.  Thus in this case we have to deal only with
the matrix $L^{(N)}\/$, and the conditions $L^{(N)}\geq 0\/$
for all $N\geq 0\/$ are both necessary and sufficient for
positivity of the underlying Hamburger distribution.

It is immediate to relate the minimal local conditions
(3.3,4) to the above theorem.  Nonnegativity of
$L^{(N)},\tilde{L}^{(N)}\/$ demands as a necessary condition
nonnegativity of every $2\times 2\/$ determinant obtained
from $2\times 2\/$ blocks along the principal diagonals of
$L^{(N)}\/$ and $\tilde{L}^{(N)}$.  This is precisely the
condition (3.3).  It is also clear why these minimal
conditions are only necessary: positivity of the diagonal
$2\times 2\/$ blocks of $L^{(N)}, \tilde{L}^{(N)}\/$ does not
capture in its entirety the positivity of $L^{(N)}\/$ and
$\tilde{L}^{(N)}$.

We now go beyond the minimal conditions (3.3,4) and
derive the next hierarchy of local conditions for a
classical PND.  Given $\{q_n\}\/$ we define
\begin{eqnarray}
x_n = q_{n-1} q_{n+1}/q^2_n\;,\;n=1,2,\ldots .
\end{eqnarray}
This definition is legitimate since in any case for a
classical PND each $q_n > 0$.  Then the minimal conditions
(3.3) simply read: 
\begin{eqnarray}
 \mbox{Classical
PND}\Rightarrow x_n\geq 1\;,\;n=1,2\ldots .  
\end{eqnarray}
As one may anticipate, the next order local conditions
which we now derive, and which are still only necessary
conditions for classicality, involve five consecutive
$q_n$'s, or equivalently three consecutive $x_n$'s.  A
necessary condition for the nonnegativity of
$L^{(N)},\tilde{L}^{(N)}\/$ is that their diagonal $3\times
3\/$ blocks be nonnegative.  That is, 
\begin{eqnarray}
 A_n =
\left(\begin{array}{ccc}
q_{n-2}&q_{n-1}&q_n\\q_{n-1}&q_n&q_{n+1}\\q_n&q_{n+1}&q_{n+2}
\end{array}\right) \geq 0\;,\;n=2,3,\ldots .
\end{eqnarray}
If we bring in $x_{n-1},x_{n},x_{n+1}\/$ here and eliminate
$q_{n\pm2}\/$ in favour of the other variables, this reads:
\begin{eqnarray}
A_n &=& \frac{1}{q_{n}}\left(\begin{array}{ccc}
q^2_{n-1}x_{n-1}&q_{n-1}q_n&q^2_n\\
q_{n-1}q_n&q^2_n&q_nq_{n+1}\\
q^2_n&q_nq_{n+1}&q^2_{n+1}x_{n+1} \end{array}\right)\geq
0\;,
\nonumber \\
n&=&2,3,\ldots .  
\end{eqnarray}
This entails three conditions: $q^2_{n-1} x_{n-1}\geq
0,\;x_{n-1}-1\geq 0,\;\det A_n\geq 0\/$.  The first two are
already obeyed; and after elementary algebra the third
reduces to a condition expressible in terms of three
consecutive $x_n$'s: 
\begin{eqnarray}
(x_{n-1}-1)(x_{n+1}-1) \geq \left(\frac{\displaystyle
x_{n}-1} {\displaystyle x_{n}}\right) ^2\;,\; n=2,3,\ldots
.
\end{eqnarray}
These are our next to minimal or {\em second order local
conditions} on the PND for classicality.

To conclude this Section, we present an interesting
implication of these conditions.  As already noted, if
$\{p_n\}\/$ is Poissonian then $\{q_n\}\/$ is a geometric
sequence, so $x_n=1\/$ identically and (3.3) are saturated.
We now ask whether it is possible to have a classical state
for which $q_{n-1} q_{n+1} = q_n^2\/$ for some values of
$n\geq 1$, whereas $q_{n-1} q_{n+1} > q^2_n\/$ for other
$n$'s.  Such classical states, if they exist, can be said
to be {\em locally Poissonian} at the former values of $n\/$,
namely whenever $x_n=1$.

Suppose a classical state is locally Poissonian at $n_0\/$,
with $x_{n_{0}}=1$.  Then two applications of (4.13), once
with $n=n_0-1\/$ and then with $n=n_0+1\/$, lead to
$x_{n_{0}\pm 1}=1$.  Continuing this process we end up with
$x_n=1\/$ for all $n\geq 1$.  Thus a classical state cannot
be only locally Poissonian : it is either locally
Poissonian throughout, $(x_n=1\/$ for all $n\geq 1\/$) or
locally superpoissonian throughout ($x_n>1\/$ for all $n$).
Translating this to $\{p_n\}\/$ we can now strengthen our
minimal conditions (3.4) to say:

For a classical PND, either 
\begin{eqnarray}
 p_{n-1}
p_{n+1} &=& \frac{n}{n+1} p^2_n
\;(\mbox{Poissonian}),\;n\geq 1
\nonumber\\ &\mbox{or}&\nonumber\\ p_{n-1} p_{n+1} &>&
\frac{n}{n+1} p^2_n \;(\mbox{Superpoissonian}),
\;n\geq 1.  
\end{eqnarray}
This is a refinement of the minimal or first order
necessary local classicality conditions (3.4) in the light
of the second order conditions (4.13).

It is now clear that the constant phases allowed in
patterns (i) and (ii) under (g) preceding eqn.(2.31) are
forbidden by the above refined condition (4.14).  This
renders Fig.2 generic and exhaustive.

\section{Alternative formulation via factorial moments}
\setcounter{equation}{0}

In this Section we present an approach to classicality of a
PND based on the normal ordered moments of $\hat{\rho}\/$ or
factorial moments of $\{p_n\}\/$, namely the sequence
$\{\gamma_m\}\/$ defined in eqn.(2.17).  This approach will
be along the lines of earlier work of Agarwal and
Tara~\cite{agarwal-pra-1992}; however our conditions for
classicality will be both {\em necessary and sufficient}.
As is amply clear from the discussion in Section~II,
though, we must bear in mind that for the vast majority of
states $\hat{\rho}\/$ and PND's $\{p_n\}\/$, the $\gamma_m\/$ are
not all defined.

Suppose we have a PND $\{p_n\}\/$ whose factorial moments
$\{\gamma_m\}\/$ are all finite and known.  (This certainly
places very severe restrictions on the behaviour of $p_n\/$
for large $n$).  Our problem is to find necessary and
sufficient conditions on $\{\gamma_m\}\/$ in order that the
PND be classical.  Now the $\gamma_m\/$ are the moments of
${\cal P}(I)\/$, just as in the preceding Section the $q_n\/$
were moments of $\tilde{\cal P}(I)={\cal P}(I)e^{-I}$.  At
the risk of repetition, the finiteness of all $\gamma_m\/$
places very strong restrictions on ${\cal P}(I)$.

It is now clear that we have again a Stieltjes moment
problem.  We want the necessary and sufficient conditions
on $\{\gamma_m\}\/$ to ensure pointwise nonnegativity of
${\cal P}(I)$.  Form two matrices $M^{(N)},\tilde{M}^{(N)}\/$
of dimension $(N+1)\/$ using the $\gamma$'s: 
\begin{eqnarray}
M^{(N)}&=& \left(\begin{array}{ccccc}
\gamma_0&\gamma_1&\gamma_2&\ldots&\gamma_N\\
\gamma_1&\gamma_2&\gamma_3&\ldots&\gamma_{N+1}\\ \vdots\\
\gamma_N&\gamma_{N+1}&\gamma_{N+2}&\ldots&\gamma_{2N}
\end{array}\right)\;;\nonumber\\ \tilde{M}^{(N)}&=&
\left(\begin{array}{ccccc}
\gamma_1&\gamma_2&\ldots&\ldots&\gamma_{N+1}\\
\gamma_2&\gamma_3&\ldots&\ldots&\gamma_{N+2}\\
\vdots\\
\gamma_{N+1}&\gamma_{N+2}&\ldots&\ldots&\gamma_{2N+1}
\end{array}\right).  
\end{eqnarray}
 Both are real
symmetric; $\tilde{M}^{(N)}\/$ equals $M^{(N+1)}\/$ with first
column and last row deleted; and
$M^{(N)}\left(\tilde{M}^{(N)}\right)\/$ has $\gamma_m\/$ for
$m\/$ even (odd) along the diagonal.  Thus, exactly as in
Section~IV, we have:

\leftline {\em \bf Theorem 2:}
The necessary and sufficient conditions for the PND
$\{p_n\}\/$ to be classical (given the existence of all
factorial moments $\gamma_m\/$) are 
\begin{eqnarray}
M^{(N)}\geq 0,\;\tilde{M}^{(N)}\geq 0,\;N=0,1,2,\ldots
\end{eqnarray}
The proof is exactly parallel to that in Section~IV, with
$\tilde{\cal P}(I),\;q_n,\;L^{(N)},\;\tilde{L}^{(N)}\/$
replaced respectively by ${\cal
P}(I),\;\gamma_n,\;M^{(N)},\;\tilde{M}^{(N)}\/$.

This theorem completes the work initiated by Agarwal and
Tara~\cite{agarwal-pra-1992}, by improving their necessary
conditions for classicality (they had only the condition
$M^{(N)}\geq 0\/$) into the necessary and sufficient
condition~(5.2).  Thus the constraints on the factorial
moments arising from the requirements $M^{(N)}\geq 0\/$ are
the same as in their work: with $N=1\/$ we have
$\gamma_0\gamma_2-\gamma^2_1\geq 0\/$, ie.,
$\gamma_2-\gamma^2_1\geq 0\/$ which (see eqn.(2.18)) is the
same as requiring the Mandel Q-parameter to be nonnegative.
With $N=2\/$ we obtain the additional condition $\det
M^{(2)}\geq 0\/$, and so on as in~\cite{agarwal-pra-1992}.
However the constraints arising from $\tilde{M}^{(N)}\geq
0\/$ are new.  For $N=0\/$ we have $\gamma_1\geq 0\/$; for $N=1\/$
we have $\gamma_1\gamma_3-\gamma_2^2\geq 0\/$; and so on.

To conclude this Section, let us re-examine the class of
pure states $|\psi(z_0,\theta)>\/$ defined in eqn.(2.32)
within the present approach.  All the $\gamma_m\/$ do exist
in this case and are easily calculated.  
\begin{eqnarray}
\gamma_m &=&
\langle \psi(z_0,\theta)|\hat{a}^{\dag m}\hat{a}^m|\psi(z_0,\theta)\rangle 
\nonumber\\ &=&|z_0|^{2m}\frac{\left(1+(-1)^m
e^{-2|z_{0}|^{2}}\cos \theta\right)}
{\left((1+e^{-2|z_{0}|^{2}}\cos\theta\right)} .
\end{eqnarray}
This leads to the following matrices $M^{(N)}\/$ and
$\tilde{M}^{(N)}\/$: 
\begin{eqnarray}
M^{(N)}&=&A\left(\begin{array}{ccccc}
1&\sigma&1&\sigma&\dots\\ \sigma&1&\sigma&1&\ldots\\
1&\sigma&1&\sigma&\ldots\\
\ldots&\ldots&\ldots&\ldots&\ldots
\end{array}\right)\;A\;,\nonumber\\
\tilde{M}^{(N)}&=&B\left(\begin{array}{ccccc}
\sigma&1&\sigma&1&\ldots\\ 1&\sigma&1&\sigma&\ldots\\
\sigma&1&\sigma&1&\ldots\\
\ldots&\ldots&\ldots&\ldots&\ldots
\end{array}\right)\;B\;,\nonumber\\
A&=&\mbox{diag}\left(1,|z_0|^2,|z_0|^4,|z_0|^6,\ldots\right)\;,\nonumber\\
B&=&\mbox{diag}\left(|z_0|,|z_0|^3,|z_0|^5,\ldots\right)\nonumber\\
\sigma&=&\frac{\left(1 - e^{-2|z_{0}|^{2}}\cos\theta\right)}{
\left(1+e^{-2|z_{0}|^{2}}\cos\theta\right)}\; .
\end{eqnarray}
It is clear that both $M^{(N)}\/$ and $\tilde{M}^{(N)}\/$ are
matrices of rank 2.  It is further clear from their
structures that $M^{(N)}\geq 0\/$ if and only if $\sigma\leq
1\/$, while $\tilde{M}^{(N)}\geq 0\/$ if and only if
$\sigma\geq 1\/$.  We thus get again the same results as in
Section~II: for $\theta\neq\pm\pi/2\/$, the state
$|\psi(z_{0},\theta)\rangle \/$ has a nonclassical PND.

We wish to remark once again that the fact that the
conditions $\tilde{M}^{(N)}\geq 0\/$ in (5.2) are needed over
and above $M^{(N)}\geq 0\/$ to form a set of necessary and
sufficient conditions for classicality is ultimately due to
the fact that our moment problem is a Stieltjes problem:
had it been a Hamburger problem, the Agarwal-Tara conditions
$M^{(N)}\geq 0\/$ would have been both necessary and
sufficient!

\section{Connection between the two approaches}
\setcounter{equation}{0}

We have presented two approaches to the problem of phase -
insensitive nonclassicality of a quantum state
$\hat{\rho}\/$: one based on $\{q_n\}\/$, the moment sequence
of $\tilde{{\cal P}}(I)\/$; the other dual approach based on
$\{\gamma_m\}\/$, the moment sequence of ${\cal P}(I)\/$ when
defined.  In each case we obtained necessary and sufficient
conditions for a PND to be classical, exploiting the fact
that the underlying problem was a Stieltjes moment problem.
In this Section we bring out explicitly the connection
between these dual approaches and establish their
equivalence, in the case (naturally) when all $\gamma_m\/$
are defined.

The fact that $\tilde{\cal P}(I)={\cal P}(I)e^{-I}\/$
suggests the use of Laplace transforms~\cite{widder-book}.
Let $\Phi(s),\tilde{\Phi}(s)\/$ be the Laplace transforms
of ${\cal P}(I),\tilde{\cal P}(I)\/$ respectively: 
\begin{eqnarray}
 \Phi(s)
&=&\int\limits^{\infty}_{0} dI\; {\cal P}(I)\; e^{-sI}\;,
\nonumber\\
\tilde{\Phi}(s)&=&\int\limits^{\infty}_{0} dI\;\tilde{\cal
P}(I)\; e^{-sI}\nonumber\\ &=& \Phi(s+1)\; .
\end{eqnarray}
We now exploit the fact that the moments of a distribution
are simply related to the derivatives of its Laplace
transform at the origin: 
\begin{eqnarray}
\gamma_n=\int\limits^{\infty}_{0} dI\;{\cal P}(I)\;
I^n &=& (-1)^n\frac{\displaystyle d^n\Phi(s)}
{\displaystyle ds^n}\big|_{s=0}\;;\nonumber\\
q_n=\int\limits^{\infty}_{0} dI\;{\cal P}(I)\; e^{-I}\;
I^n&=&(-1)^n\frac{\displaystyle
d^{n}\tilde{\Phi}(s)} {\displaystyle
ds^{n}}\big|_{s=0}\nonumber\\ &=&(-1)^n\frac{\displaystyle
d^n\Phi(s)} {\displaystyle ds^n}\big|_{s=1}\;.
\end{eqnarray}
Making two Taylor series expansions of $\Phi(s)\/$, once
about $s=0\/$ and then about $s=1\/$, equating the two
expansions and using eqn.(6.2), we have: 
\begin{eqnarray}
\sum\limits^{\infty}_{k=0} (-1)^k \gamma_k s^k/k! =
\sum\limits^{\infty}_{\ell=0}(-1)^{\ell} q_{\ell}
(s-1)^{\ell}/\ell!\;\;.  
\end{eqnarray}
Equating derivatives of both sides first at $s=0\/$ and then
at $s=1\/$ gives the pair of relations 
\begin{eqnarray}
\gamma_n&=&\sum\limits^{\infty}_{k=0} q_{n+k}/k!\;,\;n\geq 0\;;
\nonumber\\ q_n&=&\sum\limits^{\infty}_{k=0} (-1)^k
\gamma_{n+k}/k!\;,\; n\geq 0\;.
\end{eqnarray}
Writing the two sequences $\{q_n\}, \{\gamma_n\}\/$ as two
column vectors $Q\/$ and $\Gamma\/$ respectively we have the
infinite matrix equations 
\begin{eqnarray}
 \Gamma&=&
S\;Q\;,\nonumber\\ Q&=& S^{-1}\;\Gamma\;,\nonumber\\
S&=&\left(\begin{array}{ccccc} 1&1/1!&1/2!&1/3!&\ldots\\
0&1&1/1!&1/2!&\ldots\\ 0&0&1&1/1!&\ldots\\
\ldots&\ldots&\ldots&\ldots&\ldots
\end{array}\right)\;,\nonumber\\
S^{-1}&=&\left(\begin{array}{ccccc}
1&-1/1!&1/2!&-1/3!&\ldots\\ 0&1&-1/1!&1/2!&\ldots\\
0&0&1&-1/1!&\ldots\\ \ldots&\ldots&\ldots&\ldots&\ldots
\end{array}\right)\;.
\end{eqnarray}
We have displayed $S,S^{-1}\/$ to exhibit the fact that these
matrices are unchanged if the first row and first column
are deleted.  We shall have occasion to return to this
important feature.

While $Q\/$ and $\Gamma\/$ considered as (infinite dimensional)
column vectors are connected by the matrix $S\/$, the
corresponding infinite dimensional matrices
$L=L^{(\infty)}\/$ and $M=M^{(\infty)}\/$ are connected through
a symmetric transformation using $S^{1/2}\/$.  We have:
\begin{eqnarray}
(S^{1/2})_{jk}&=& 2^{j-k}/(k-j)!\;,\nonumber\\
(S^{-1/2})_{jk}&=& (-2)^{j-k}/(k-j)!\;\;.  
\end{eqnarray}
(So both matrices are upper triangular: these matrix
elements vanish for $k<j\/$).  That $S^{1/2}\/$ and $S^{-1/2}\/$
so defined are indeed inverses of one another follows
simply from the familiar properties of binomial
coefficients.  Using these same properties it may be
verified that 
\begin{eqnarray}
 M&=&S^{1/2}
L(S^{1/2})^T\;,\nonumber\\ L&=&S^{-1/2}M(S^{-1/2})^T\;.
\end{eqnarray}
This proves that $M\geq 0\/$ if and only if $L\geq 0\/$.

Let $\tilde{Q}\/$ be the moment sequence derived from $Q\/$ by
simply dropping $q_0\/$.  Clearly,
$\tilde{L}=\tilde{L}^{(\infty)}\/$ is in the same relation to
$\tilde{Q}\/$ as $L\/$ is to $Q\/$. Similarly if we form
$\tilde{\Gamma}\/$ from $\Gamma\/$ by simply dropping
$\gamma_0\/$, then $\tilde{M}=\tilde{M}^{(\infty)}\/$ is in the
same relation to $\tilde{\Gamma}\/$ as $M\/$ is to $\Gamma\/$.
Now recalling the fact that $S,S^{-1},S^{1/2},S^{-1/2}\/$
have the interesting property of each being unchanged upon
deletion of the first row and first column we conclude:
\begin{eqnarray}
\tilde{\Gamma} &=& S
\tilde{Q}\;,\;\tilde{Q}=S^{-1}\tilde{\Gamma}\;;
\nonumber\\ \tilde{M}&=&S^{1/2}\tilde{L}(S^{1/2})^T\;,\;
\tilde{L} = S^{-1/2} \tilde{M}(S^{-1/2})^T\;.
\end{eqnarray}
This proves that $\tilde{M}\geq 0\/$ if and only if
$\tilde{L}\geq 0\/$.  Combined with the statement after
eqn.(6.7), we have thus established the equivalence of the
two approaches. They are dual to one another, and in
particular Theorem 1 is equivalent to Theorem 2.

The above analysis shows that if we drop the first $k\/$
terms from the moment sequence of a bonafide Stieltjes
probability distribution (in the semi-infinite interval),
the result is again a bonafide Stieltjes moment sequence
(of some other valid probability distribution).  This is a
distinguishing feature of the Stieltjes moment problem.  It
is not difficult to see that the corresponding statement is
true for the Hamburger moment problem (on the entire real
line) only if an {\em even} number of initial terms are
dropped from the moment sequence.

\section{Concluding remarks}
\setcounter{equation}{0}

Every probability distribution $\{p_n\}\/$ over the
nonnegative integers, admissible according to the laws of
classical probability theory, can appear as the PND of some
quantum state $\hat{\rho}\/$ of the single mode radiation
field.  We have given explicit necessary and sufficient
conditions - in two equivalent or dual forms - for a PND to
be the result of a classical state in the sense appropriate
to quantum optics.  From the perspective of classical
probability theory, the significant points are these: that
we are able to view the quantities $q_n=n! p_n\/$ as the
moments, in the Stieltjes sense, of an auxiliary function
$\tilde{\cal P}(I)\/$, and that the properties of
$\tilde{\cal P}(I)\/$ or equally well of ${\cal P}(I)\/$
determine whether the state is classical or not.  The route
from $\{p_n\}\/$ to ${\cal P}(I)\/$ is via the pair of
eqns.(2.14,15).

We have emphasized the following points in our work: (i)
that local classicality conditions on the PND are naturally
available for all states $\hat{\rho}\/$; and (ii) all
statements based on the factorial moments make sense only
for a quite limited set of states $\hat{\rho}\/$, for which
the probabilities $p_n\/$ go to zero ``sufficiently fast'' as
$n\rightarrow\infty\/$.  It seems to us that some of these
points have not been given in the past as much attention as they
deserve.  Above this, we stress the
completeness of our results, and the more careful treatment
of oscillations in the PND as a signature of
nonclassicality, which becomes possible with our local
approach.

\end{document}